\magnification1200

\def\today{\number\year\space \ifcase\month\or 	January\or February\or 
	March\or April\or May\or June\or July\or August\or September\or
	October\or November\or December\fi\space \number\day}


\font\bigbf=cmbx12


\def\degrees{\hbox{${}^\circ$\hskip-3pt .}}
\def\pp{\par\hangindent=.125truein \hangafter=1}
\def\aref#1;#2;#3;#4{\pp #1, {\it #2}, {\bf #3}, #4}
\def\abook#1;#2;#3{\pp #1, {\it #2}, #3}
\def\arep#1;#2;#3{\pp #1, #2, #3}
\def\spose#1{\hbox to 0pt{#1\hss}}
\def\simlt{\mathrel{\spose{\lower 3pt\hbox{$\mathchar"218$}}
     \raise 2.0pt\hbox{$\mathchar"13C$}}}
\def\simgt{\mathrel{\spose{\lower 3pt\hbox{$\mathchar"218$}}
     \raise 2.0pt\hbox{$\mathchar"13E$}}}

\def\frac#1/#2{\leavevmode\kern.1em
 \raise.5ex\hbox{\the\scriptfont0 #1}\kern-.1em
 /\kern-.15em\lower.25ex\hbox{\the\scriptfont0 #2}}

\hyphenation{Nity-an-anda}

\def\gum{$\gamma$UMi}

\line{\hfil CfPA-94-TH-54}
\line{\hfil astro-ph/9410088}
\line{\hfil \today}
\bigskip

\centerline{\bigbf A First Map of the CMB at $0\degrees5$ Resolution}
\bigskip

\centerline{Martin White and Emory F. Bunn}
\smallskip

\centerline{\it Center for Particle Astrophysics and Departments of
Astronomy \& Physics}
\centerline{\it University of California, Berkeley, CA 94720-7304}

\bigskip
\bigskip

\centerline{ABSTRACT}

\noindent
We use a Maximum Entropy technique to reconstruct a map of the microwave sky
near the star Gamma Ursae Minoris, based on data from flights 2, 3 and 4
of the Millimeter-wave Anisotropy eXperiment (MAX).
\medskip

{\it Subject headings:} cosmology: cosmic background radiation,
numerical methods


\bigskip
\noindent {\bf Introduction}
\medskip

The data from cosmic microwave background (CMB) anisotropy experiments is
improving rapidly.  The next generation of degree--scale experiments should
have the capability to map significant regions of the sky.
However, due to uneven sky coverage, differencing strategies, and noise in
the data, reconstructing a temperature map from the observations will be
non--trivial.  It is therefore of interest to begin developing techniques
for this task.

The Millimeter-wave Anisotropy eXperiment (MAX) has now accumulated data
on the temperature anisotropies in the CMB near the star Gamma Ursae Minoris
(\gum) from 3 separate flights.
(Alsop et al.~1992, Meinhold et al.~1993, Devlin et al.~1994).
This data covers $\sim 10^{\circ}\times5^{\circ}$ on the sky, with the central
$5^{\circ}\times2^{\circ}$ being densely sampled (see Fig.~1), and thus is
ideal for testing algorithms for constructing maps of the microwave sky.

In studies of CMB anisotropies it has been common to assume that the
fluctuations are gaussian distributed, and that the overall phases of the
anisotropy in {\it our} universe are irrelevant.
In this paper we would like to consider an alternative approach, in which
maps of the microwave sky in particular regions are made.
As an example of the method, we concentrate on the region of sky near
\gum\ where a large data set already exists.
The advantage of making CMB maps, apart from the simple desire to map the
whole sky in as many wavebands as possible, is that it allows us to develop
a catalog of features for comparison with other experiments.
In addition one can look for properties of the sky which are not predicted by
theories, and could be overlooked in statistical analyses.
While detailed comparison of observations of the same region of the sky by
different experiments should be done by statistical comparison of the raw data
sets (e.g.~cross-correlation of the temperature difference maps in current
experiments), having a map of the sky in the region of interest is helpful in
planning flights and obtaining a visual representation of the region to be
surveyed.
Additionally these maps can be used to correlate phase information through the
method of constrained realizations
(e.g.~Hoffman \& Ribak~1991, Bunn et al.~1994)
which can allow predictions to be made for other experiments.

Any inversion procedure of this kind, where we attempt to reconstruct a
temperature map from a small region of the sky where only temperature
differences are measured, requires a regularization procedure.
For example, to construct a map one needs information on the long wavelength
contributions which are not well constrained by the data set.
In this paper we shall adopt the {\it maximum entropy} procedure (MaxEnt),
often used in other branches of astronomy (see below).
MaxEnt provides a method for choosing among the many maps which could lead
to the observed data.
The advantage of this method for our purposes is that it reconstructs the
``smoothest'' maps consistent with the data.
If it is our intent to search for features in the maps, such as hot or cold
spots or indications of non-gaussian structures (such as lines), this is
clearly the most conservative regularization.
Whenever the inversion is not unique a choice needs to be made, and we take
the stance that we wish to introduce only those features which are
{\it required} by the data, even if this can miss features which are
{\it allowed} by the data.

\bigskip
\noindent {\bf Maximum Entropy Method}
\medskip

The main use of the maximum entropy method is to provide a regularized
inversion procedure for noisy and incomplete data.
In practice what one does is construct a model, which in our case consists
of the temperature values in $N_{\rm pix}=64\times64$ pixels on a
$30^{\circ}\times8^{\circ}$ patch of the sky.
We attempt to find the best model sky given data $T_k\pm\sigma_k$ at some
locations $\hat{n}_k$ with $k=1\ldots N_{\rm obs}$.
We assume the errors, $\sigma_k$, are gaussian and independent.
In the MaxEnt method we minimize not the $\chi^2$ associated with the fit
of this ``theory'' to the data, but rather the combination
$$ S + \lambda\,\chi^2 \eqno(1)
$$
where the entropy $S$ will be discussed in detail below and $\lambda$ is a
Lagrange multiplier which determines the relative weight given to the two
terms.
The first term ($S$) will be extremized when the map is as featureless
as possible, while the second tries to make the model temperature values
agree with the data as closely as possible.  Maximizing the combination
should lead to the smoothest map consistent with the data, and has the
advantage of being computationally simple.
Our result will be the most conservative picture of deviations from uniformity
consistent with the data.

There is a large body of literature on the MaxEnt method.  For information
on its use in an astronomical context see
e.g.~Gull \& Daniel~(1978), Burch, Gull \& Skilling~(1983),
Cornwell \& Evans~(1984), Naryan \& Nityananda~(1986),
Lahav \& Gull~(1989), Skilling~(1991), Press et al.~(1992),
or the proceedings of the several MaxEnt workshops and conferences
(and the references therein).

Traditionally, when dealing with intensity maps, one takes the entropy
function to be $S =-I\log I$ which requires $I\ge 0$.
In our case the analogous procedure would be to use as our model the absolute
temperature on the sky (obtained by requiring the data have an average
temperature of $2.726$K for example).
However, since the temperature fluctuations in the data are 1 part in $10^5$
this can lead to numerical problems.
Using simply the temperature differences as our model (i.e.~subtracting a
mean $T$ from each point) leads to problems with the evaluation of the
logarithm in $S$.
Note however that the role of the entropy function is to ``smooth'' the
inversion.  The overall offset and range of the pixel values in the model
is not relevant to this question.
Mathematically we can say the extremum of the entropy is invariant under
rescalings and shifts of the origin, so we rescale the temperatures before
computing the logarithm.

There are other definitions of the entropy which can be used
(see Naryan \& Nityananda~(1986) for a discussion).
In the context of Bayesian inference, a choice of $S$ corresponds to a choice
of prior information, and the inferences should be robust under changes in
``prior''.  We shall stick with the $I\log I$ definition since it enforces
``smooth'' maps.

This leads us to consider the relative weight assigned to the entropy and
the $\chi^2$, which is given by $\lambda$.
Often one assigns to $\lambda$ the value required so that the $\chi^2$ of
the best fit model is approximately equal to the number of data points.
(In this case $\lambda$ is a Lagrange multiplier introduced to enforce the
$\chi^2$ constraint while maximizing the entropy.)
This can be implemented straightforwardly during the solution of the MaxEnt
inversion.
Alternatively (Gull~1989, Lahav \& Gull~1989) one can use the eigenvalues of
the Hessian matrix, which is less arbitrary since it can be justified from
Bayesian considerations.
Generally the resulting map is independent of the choice of $\lambda$ over
a large range, and hence we will stick with the former prescription even though
it has an element of arbitrariness.

Having fixed $\lambda$ one proceeds to solve the data by an iterative
method.  A more generic and powerful alternative is a maximum search
algorithm (Burch et al.~1983, Cornwell \& Evans~1985); however, we have
found the iterative solution to converge stably and rapidly to the desired
solution, obviating the need for complicated strategies.
Let us call $t_j$ the temperature values in our ``model'' map, where $j$
runs from 1 to $N_{\rm pix}$.
Let $B_{k,j}$ be the beam profile function which acts on the model pixels
and predicts the measured temperature differences
$\tau_k=\sum_j t_j B_{k,j}$ ($k=1\ldots N_{\rm obs}$).
One now starts with a uniform map and iterates the equation
(Gull \& Daniell~1978)
$$
t_j = \exp\left[-{\rm const}_1+\lambda \sum_{k=1}^{N_{\rm obs}}
  B_{k,j} {T_k-\tau_k\over\sigma_k^2} \right] + {\rm const}_2
\eqno(2)
$$
where recall $T_k\pm\sigma_k$ are the temperature differences to be fit.
This equation comes from extremizing Eq.(1) with respect to the $t_j$, and the
values of the constants specify the zero and range of the temperature scale.
We have chosen these constants so that unmeasured parts of the map are
assigned zero temperature offset.

\bigskip
\noindent {\bf MAX Data Set}
\medskip

The MAX experiment is a sub--degree scale CMB anisotropy experiment which
measures temperature differences on the sky over a range of frequencies.
In the following we shall assume that all of the signal seen in the \gum\ 
scan is purely that of CMB anisotropies, so we co-add all the frequency
channels to increase the signal-to-noise ratio.  If this turns out not to be
the case, the method could also be implemented for each frequency separately.

The temperature ``differences'' are defined by performing a sinusoidal chop,
at a frequency of $\nu=6\,$Hz, with amplitude $\alpha_0=0\degrees65$ parallel
to the scan direction of the telescope.  The signal is demodulated at the first
harmonic of the chop frequency so that the temperature assigned to the point
$\hat{n}$ is
$$
\widetilde{T}(\hat{n}) \equiv -3.34\,\int_0^{1/\nu} dt\ \nu \sin(2\pi\nu t)\ 
\Theta\left(\hat{n}\cos\alpha(t) + \hat{i}\sin\alpha(t) \right)
\eqno(3)
$$
where $\alpha(t)=\alpha_0\sin(2\pi\nu t)$, $\hat{i}$ is a unit vector lying
along the chop direction and perpendicular to $\hat{n}$, and $\Theta(\hat{n})$
is the beam--smoothed temperature at $\hat{n}$.
See Srednicki et al.~(1993) for more details.
To obtain an effective beam profile we replace $\Theta$ in the above with the
beam weight function, which we take to be a gaussian of width $0\degrees5$FWHM.
The beam width varied slightly from flight to flight, which we have included
in the analysis.

The data available from the MAX group has had an offset subtracted from each
`scan'.  This has the effect of modifying the long wavelength modes
reconstructed by our procedure.  Since these modes are not well constrained
by the data in any case, we have not tried to correct for this effect. 

\bigskip
\noindent {\bf Results and Conclusions}
\medskip

We show in Fig.~2a our estimated map, which is consistent (by construction)
with the 3 MAX--\gum\ data sets.
The pixels are distributed exponentially in temperature, presumably reflecting
our assumption of gaussian beams. 
The RMS of the non-zero portions of the map is consistent with the expected
fluctuations in a model such as CDM.
In Fig.~2b we show a realization of a standard CDM sky, smoothed on
$0\degrees5$ and with power on wavelengths larger than the box removed.
The maps are qualitatively similar, though there is a large degree of
arbitrariness in statistical comparison which makes it difficult to compare
quantitatively.
Our map appears to have more small--scale structure and less large--scale
structure than the CDM map, possibly reflecting that our procedure reconstructs
small--scale power better than large--scale power.
This is not unexpected given the small amount of sky ($\sim25$ square degrees)
for which we have data.
To test this we have simulated a MAX data set using our standard CDM model
and used the same algorithm to reconstruct a map based on this synthetic data.
The resulting map (not shown) is very similar to Fig.~2a, having an exponential
pixel distribution and more small--scale power than Fig.~2b.
A detailed study of the biases in this reconstruction method still remains to
be done.

In conclusion we have presented a method for constructing temperature maps
of the CMB which can be easily applied to differencing experiments which
cover only a small fraction of the sky.  As an application of this method we
have constructed the ``most likely'' picture of the microwave sky consistent
with the data in the region near \gum.
While comparison of scans and model testing should be done statistically using
the differenced data, this method allows one to check for features which are
contrary to expectations and to plan future observations.

\bigskip
\noindent {\bf Acknowledgements}
\medskip

We would like to thank Douglas Scott and Joseph Silk for useful conversations
and encouragement, and Mark Devlin and Stacy Tanaka for help with the MAX
data.

\vfill\eject

\noindent {\bf References}
\parindent0in
\frenchspacing
\medskip
\aref Alsop, D.C., et al., 1992;ApJ;395;317
\aref Bunn, E.F., et al., 1994;ApJ;432;L75
\aref Burch, S.F., Gull, S.F., \& Skilling, J., 1983;Computer Vision,
Graphics and Image Processing;23;113
\aref Cornwell, T.J., \& Evans, K.F., 1984;Astron. Astrophys.;143;77
\aref Devlin, M., et al. 1994;ApJ;430;L1
\abook Gull, S.F., 1989;{\rm in} Maximum Entropy and Bayesian Methods;ed.
J. Skilling, Kluwer Academic Publishers, Dordrecht
\aref Gull, S.F., \& Daniell, G.J., 1978;Nature;272;686
\aref Hoffman, Y., \& Ribak, E., 1991;ApJ;380;L5
\aref Lahav, O. \& Gull, S.F., 1989;MNRAS;240;753
\aref Meinhold, P.R., et al., 1993;ApJ;409;L1
\aref Narayan, R., \& Nityananda, R. 1986;Ann Rev Astron Astrophys;24;127
\abook Press, W.H., et al., 1992;Numerical Recipes; 2nd ed.,
Cambridge University Press, New York
\aref Skilling, J., 1991;Nature;353;707
\aref Srednicki, M., White, M., Scott, D. \& Bunn, E., 1993;Phys. Rev.
Lett.;71;3747
\nonfrenchspacing

\bigskip

\bigskip
\noindent {\bf Figure Captions}
\medskip

\noindent Fig.~1: The region of sky, near \gum, covered by the second, third
and fourth flights of the MAX experiment.  The solid circles represent data
from the second flight, triangles those from the third and squares those from
the fourth.
For reference we show the beam pattern as the two circles (labelled plus and
minus) for the MAX differencing strategy in the lower right.
The apparent elliptical shape is due to the axis ranges chosen for the plot.
Also shown (lower left) is the size of the pixels used in the map.

\bigskip

\noindent Fig.~2: (a) The reconstructed sky temperature, in $\mu$K.  Since
the MAX experiment measures only temperature differences, the ``zero'' of
the temperature scale is arbitrary.
(b) A sky simulated using a CDM model normalized to COBE and with $\Omega_0=1$,
$H_0=50\,{\rm km}\,{\rm s}^{-1}{\rm Mpc}^{-1}$ and $\Omega_B=0.05$, smoothed on
$0\degrees5$ and with power on wavelengths larger than the box removed.

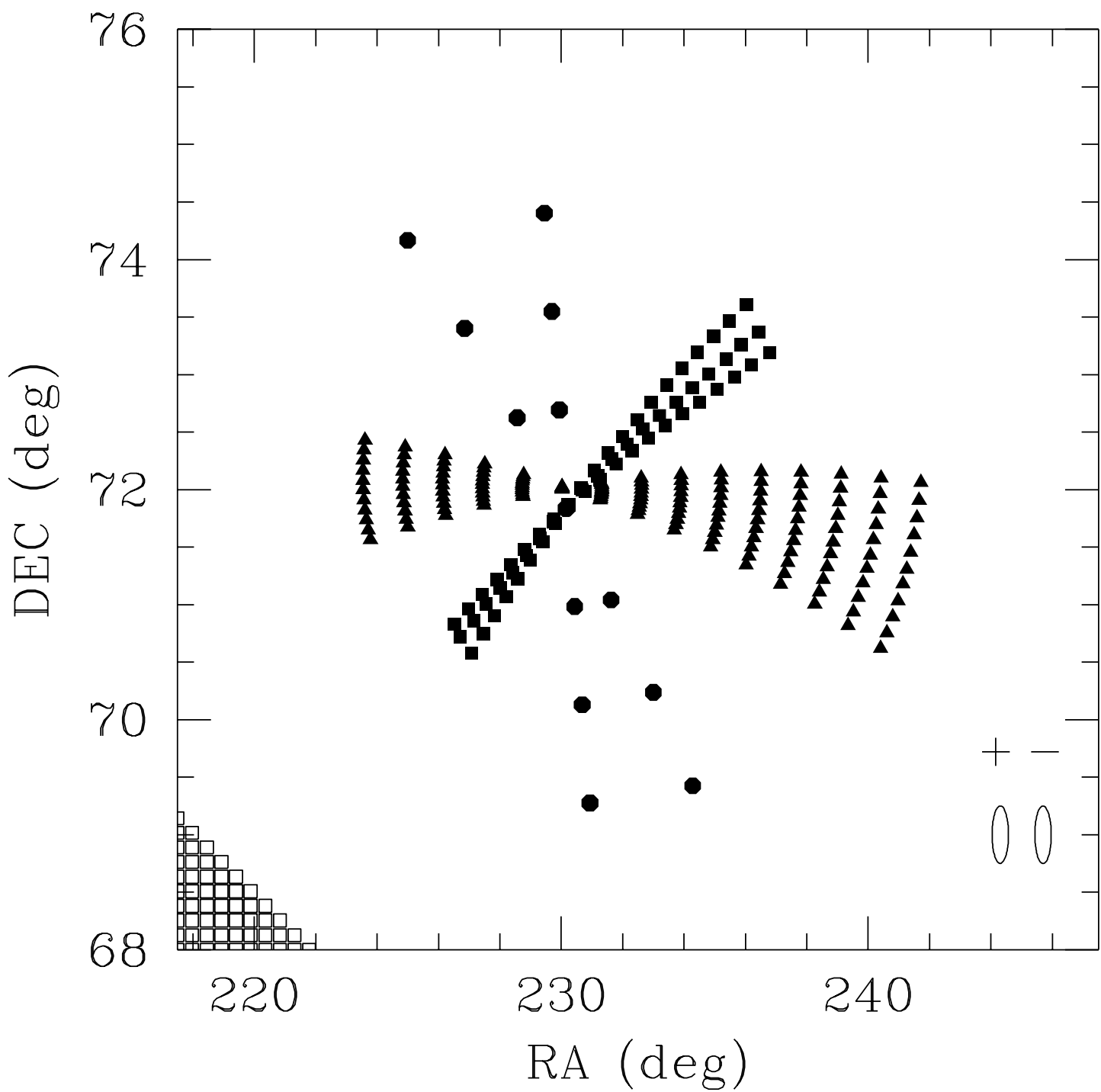
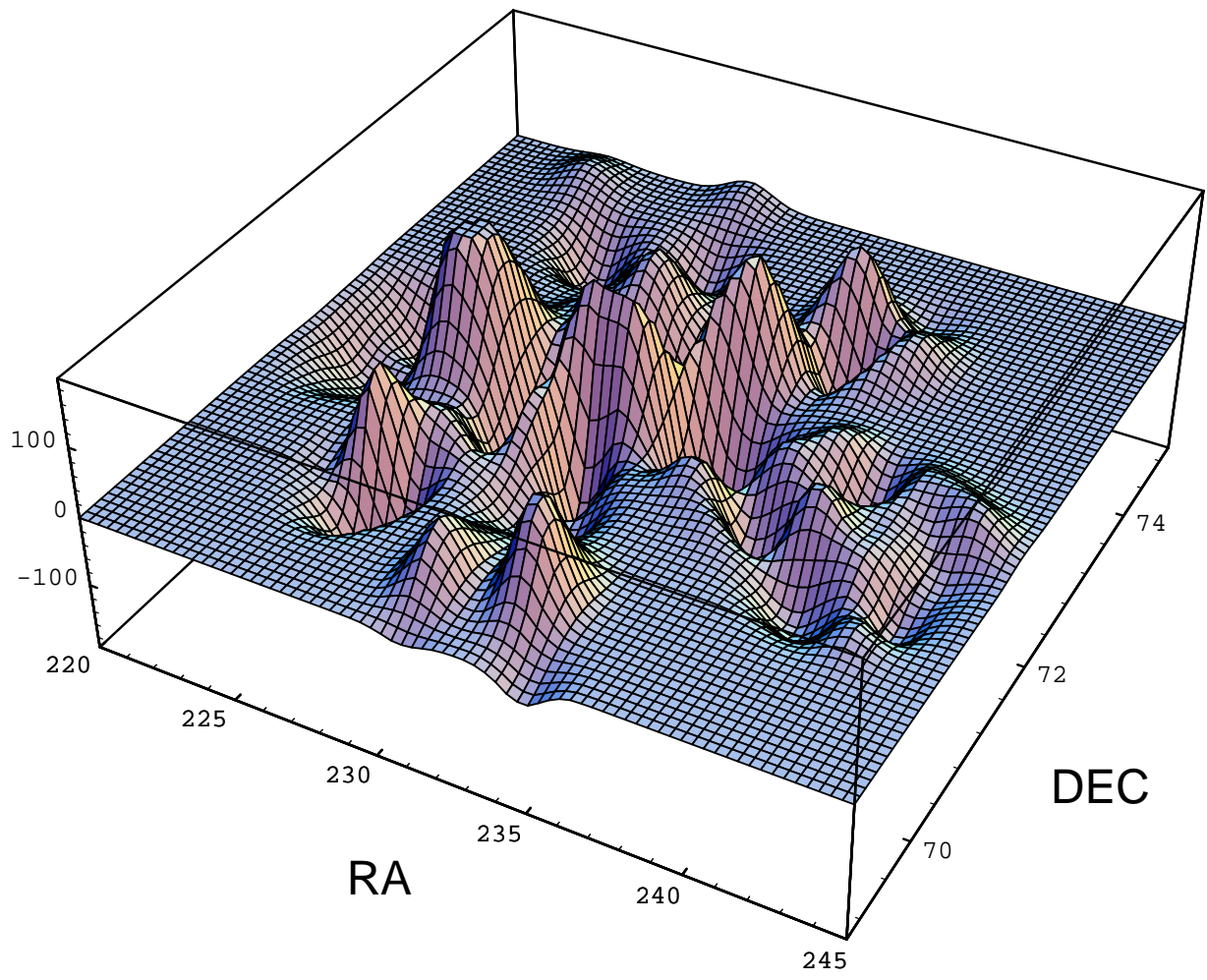
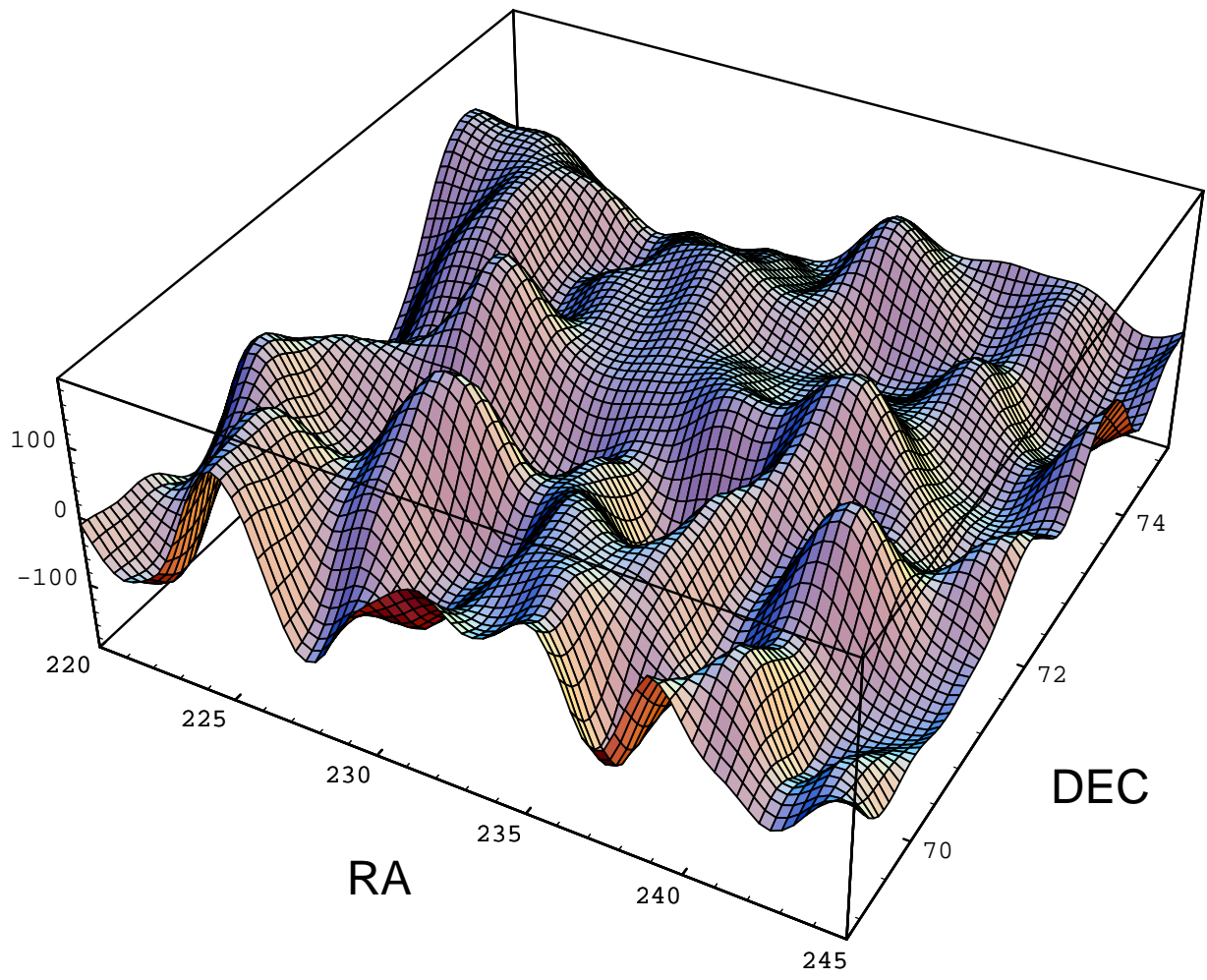
\bye